\documentclass[twocolumn,prb]{revtex4-2}
\usepackage{titlesec}
\usepackage{subfigure}
\usepackage{amsmath}
\usepackage{amsfonts}
\usepackage{amssymb}
\usepackage{feynmp}
\usepackage{subfigure,tikz}
\usepackage{graphicx}
\usepackage{setspace}
\usepackage{comment}
\usepackage{dsfont}
\usepackage{color}
\usepackage[pdftex,colorlinks=true,linkcolor=blue,citecolor=blue]{hyperref}

\begin{document}

\title{Topological doublon edge states induced by the spatially modulated interactions}

\author{Zheng-Wei Zuo}
\author{Wanwan Shi}
\author{Haisheng Li}

\affiliation{School of Physics and Engineering, Henan University of Science and Technology, Luoyang 471023, China}
\date{\today}

\begin{abstract}
 The topological properties of the one-dimensional interacting systems with spatially modulated interaction in two-particle regime are theoretically investigated. Taking the boson-Hubbard model and spinless fermion interacting model as examples, we show that the energy spectra for doublon (known as two-particle pair) as a function of modulated period exhibit the butterfly-like structure for strongly-correlated limit, whose topological features can be decoded by the topological invariants and topological nontrivial doublon bound edge states. When the nearest-neighbor hopping evolves stronger, the doublon bands could intersect with scattering bands, the one-dimensional interacting systems display the phases of topological insulators and two-particle bound states in the continuum. For a sufficiently larger nearest-neighbor hopping, the doublon collapse takes place, where both the bulk doublon states and topological doublon edge states become unstable and could dissociate into two weakly interacting bosons. For the mapped two-dimensional single-particle systems, numerical calculations manifest the existence of the topological insulator and topological metal phases with corner states located in only one or two corners.
\end{abstract}

\maketitle
\section{Introduction}
In the past two decades, the research of topological phases of quantum matter has obtained tremendous progress\cite{Bernevig13Book, MoessnerR21Book, Hasan10RMP, QiXL11RMP, Franz15RMP, ChiuCK16RMP, WenXG17RMP}. For the independent-particle systems, the topological gap phases such as topological insulators and topological superconductors, the topological metal, and topological gapless phases are the typical examples. For the many-body correlated systems, the fractional quantum Hall effect and quantum spin liquids are the prominent subjects. Nowadays, the non-Hermitian topological states\cite{BergholtzEJ21RMP}, the higher-order topological states\cite{Benalcazar17SCI, Benalcazar17PRB}, and the topological magnetic materials\cite{XiaoZY24PRX, ChenXB24PRX, JiangY24PRX} draw significant interest.

At present, in contrast to the abundant topological phases in independent-particle and many-body correlated systems, the research of topological states in the few-body interacting systems regime remains largely unexplored. Recently, the topological states in one dimensional (1D) system within the two-particle regime have begun to attract attention\cite{DiLiberto16PRA, GorlachMA17PRA, MarquesAM17PRB, MarquesAM18PRA, MarquesAM18JPCM, StepanenkoAA20PRAA, ChengDL21PRR, AzconaPM21SR, BesedinIS21PRB, NicolauE23PRA2, ShitT24arXiv, LonghiS12OL, OlekhnoNA22PRB, ZhangWX22NTC, ZhangWX23CP, KwanJ24SCI, LeeCH21PRB, FaugnoWN22PRL, ZhangWX22PRB, ShenRZ22CP, PoddubnyAN23PRB, LonghiS23PRB, BrighiP24PRA, LiuYX24PRL, MarcheA24PRA, PelegriA20PRR, ZuritaJ19AQT, KunoY20PRA, KunoY20PRB, NicolauE23PRA, ZhouXQ23PRB, OlekhnoNA20NTC, HuangBN24PRL,HuangBN24PS}. When the particle-particle interaction strengthens, the two particles could form bound pair states, namely doublon. On the other hand, as shown in Refs.\cite{Valiente08JPB, ZhangJM12PRL, ZhangJM13PRA, QinXZ14PRA}, the 1D two-particle interacting physical system can be mapped to the two dimensional (2D) independent-particle system. Thus, research on topological doublon states has been conducted across various theoretical models, including the Su-Schrieffer-Heeger (SSH) model\cite{DiLiberto16PRA, GorlachMA17PRA, MarquesAM17PRB, MarquesAM18PRA, MarquesAM18JPCM, StepanenkoAA20PRAA, ChengDL21PRR, AzconaPM21SR, BesedinIS21PRB, NicolauE23PRA2, ShitT24arXiv}, anyon systems\cite{LonghiS12OL, OlekhnoNA22PRB, ZhangWX22NTC, ZhangWX23CP, KwanJ24SCI}, non-Hermitian systems\cite{LeeCH21PRB, FaugnoWN22PRL, ZhangWX22PRB, ShenRZ22CP, PoddubnyAN23PRB, LonghiS23PRB, BrighiP24PRA, LiuYX24PRL, MarcheA24PRA, WangL25PRA, ZhangYL24arXiv, LingWZ25arXiv}, and the flat-band systems\cite{PelegriA20PRR, ZuritaJ19AQT, KunoY20PRA, KunoY20PRB, NicolauE23PRA, ZhouXQ23PRB}. Unlike the independent-particle systems, the topological edge states in the two-particle interacting systems are the two-particle bound states. When the doublon bands and the scattering bands interact, the doublon collapse appears, where both the bulk and edge doublon states become unstable and dissociate into two weakly interacting bosons. Additionally, investigations into topological doublon edge states in the 2D and three dimensional systems within the two-particle regime have emerged\cite{SalernoG18PRA, SalernoG20PRR, LinL20PRA, IskinM21PRA, BertiA22PRA, StepanenkoAA22PRL, IskinM23PRA, OkumaN23PRR, AlyurukDC24PRB}. 

The inhomogeneous particle-particle interactions such as spatially modulated types in strongly-correlated systems reveal some interesting features\cite{PaivaT96PRL, PaivaT98PRB, KogaA13JPSJ, MendesST13PRB, LiJian18PRL, ChengZY24PRB, OlekhnoNA20NTC, HuangBN24PRL,HuangBN24PS, ShavitG20PRR, ShavitG20SPP, ZuoZW20NJP, RosnerM21PRR, MondalS22PRB}. For example, the Thouless pumping of multiparticle bound states \cite{HuangBN24PRL,HuangBN24PS}, giant magnetoresistance\cite{LiJian18PRL}, fractional conductance plateaus\cite{ShavitG20PRR, ShavitG20SPP}, and these various topological states\cite{ZuoZW20NJP, RosnerM21PRR, MondalS22PRB} are uncovered. It is interesting to analyze the effect of the spatial-modulated interactions on the physical features of few-body interacting systems. In this paper, we systematically investigate the influence of the periodically spatial-modulated interactions on the topological properties of the 1D interacting spinless boson and fermion system within two-particle regime. The rest of this paper is organized as follows: In Sec. \ref{Boson}, we introduce the model of the 1D Boson-Hubbard model with pair-hopping interaction and its mapping into a 2D square lattice in two-particle regime. Concretely, Sec. \ref{OnsiteInteraction} discusses the topological features for the periodically spatial-modulated on-site interaction case. The doublon bands show a butterfly-like structure, whose topological properties can be analyzed by the Chern number. For the corresponding 2D square lattice, the system can emerge the topological metal (topological bound states in continuum) with two corner states, which locate at the two diagonal corner sites. The topological properties for the periodically spatial-modulated pair-hopping interaction case are given by Sec. \ref{PairHopping}. Sec. \ref{Fermion} analyzes the topological properties of the spinless fermion Hubbard model with pair-hopping interaction. Conclusions and discussions are reported in Sec. \ref{Conclusions}.

\section{Boson-Hubbard model case}\label{Boson}

The periodically modulated interactions have drastic effect on the topological properties of 1D interaction systems in two-particle regime. Here, we first investigate the 1D  boson-Hubbard model with a pair-hopping interaction case, whose Hamiltonian in the second quantized form is given by
\begin{align}
H_b=&\sum_j^{N-1}[-Jb_{j+1}^{\dagger}b_j+\frac{g_j}{2}b_{j+1}^{\dagger}b_{j+1}^{\dagger}b_jb_j+H.c.] \notag \\ 
&+\sum_j^N \frac{U_j}{2} n_j(n_j-1),\label{Hb}
\end{align}
in which $b_{j}^{\dagger}$ ($b_{j}$) is boson creation (annihilation) operator on the lattice site $j$ and $N$ is lattice sites. $n_j=b_j^{\dagger}b_j$ is the boson number operator and $J$ is the parameter controlling the strengths of the nearest-neighbor coupling. The second term ($g_j$) is the pair-hopping interaction from the site $j$ to the site $j+1$. The $U_j$ term parametrizes the on-site interaction at the site $j$. In this paper, we choose the pair-hopping interaction and on-site interaction as periodically spatial-modulated types. 

As shown in Refs.\cite{Valiente08JPB,ZhangJM12PRL,ZhangJM13PRA}, the two-particle physics can be mapped to the 2D single-particle regime. Now, we consider two distinguishable bosons hopping on the 1D interacting chain and take the two-particle wave function in the form
\begin{equation}
|\psi\rangle=\frac{1}{\sqrt2} \sum_{m, n} \beta_{m,n} b_m^{\dagger} b_n^{\dagger}|0\rangle \label{WaveFunction}
\end{equation}

Because of the bosonic symmetry, the $\beta_{m,n}=\beta_{n,m}$ for any indices $m$ and $n$. Inserting Eq.\ref{Hb} and Eq. \ref{WaveFunction} into the Schrödinger equation, we can obtain the linear system of equations:
\begin{align}
-J (\beta_{m-1,n}+\beta_{m+1,n} +&\beta_{m,n-1} \notag+\beta_{m,n+1})=E\beta_{m,n}, (m \neq n) \\
-2J (\beta_{m-1,n}+\beta_{m,n+1}) &+g_{m-1}\beta_{m-1,n-1}+g_m\beta_{m+1,n+1}\notag \\ 
&=(E-U_m)\beta_{m,n}, (m=n).
\end{align}

The corresponding graphic representation of the two-boson Hamiltonian in 2D is shown in Fig. \ref{FigModel}(a). In the 2D square lattice model, the on-site interaction $U_m$ and spatially modulated pair-hopping $g_m$ are represented as the on-site potential of diagonal sites and nearest-neighbor coupling between the diagonal sites. Next, we analyze the effect of the spatially modulated pair-hopping interaction and on-site interaction on the topological properties of this two-boson interacting system, respectively.

\subsection{Periodically modulated on-site interaction}\label{OnsiteInteraction}
Firstly, we set the pair-hopping interaction as uniform case $g_j=g$ and the on-site interaction as spatially periodically modulated $U_j=U\cos(2\pi\alpha j+\phi)$, where $\alpha$ is periodically modulated frequency. The two parameters $U$ and $\phi$ are the modulation strength and phase, respectively. For the experimental systems such as superconducting circuits and optical lattices of ultracold atoms, the periodically spatial-modulated on-site interaction could be tuned by controlling qubit anharmonicity of individual qubits\cite{TaoZY23arXiv} and Feshbach resonance\cite{YamazakiR10PRL}, respectively. For convenience, we temporarily neglect the nearest-neighbor coupling ($J=0$, strong-interaction limit) in the following, where the system becomes exactly solved and the two-boson states are tightly bound. Now, the Hamiltonian for the system becomes
\begin{align}
H_b=&\sum_j^{N-1}\frac{g_j}{2}[(b_{j+1}^{\dagger})^2b^2_j+H.c.] +\sum_j^N \frac{U_j}{2} n_j(n_j-1), \label{Hb1}
\end{align}
The linear equation $\beta_{m,n}$ becomes
\begin{equation}
g(\beta_{m-1,m-1}+\beta_{m+1,m+1})=(E-U_m)\beta_{m,m}. \label{diagAAH}
\end{equation}

\begin{figure}[tb]
\centering
\includegraphics[width=\columnwidth]{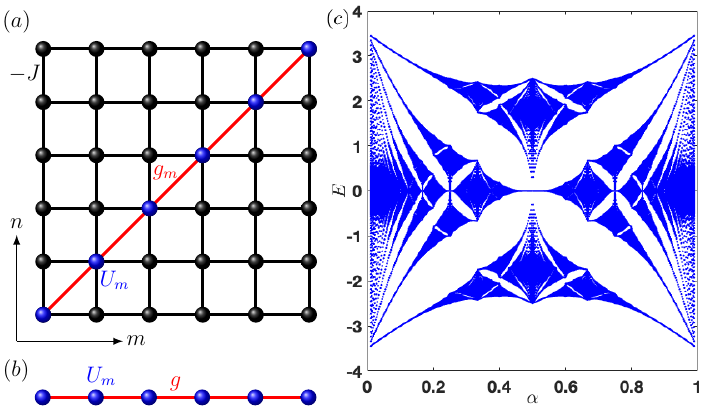}
\caption{$(a)$ Illustration of the mapping onto a 2D single-particle system. $(b)$ The effective 1D diagonal AAH model. $(c)$ The energy spectra as a function of $\alpha$ under the PBC when the system parameters are $U=1.5, \phi=0$, and $g=1$.}
\label{FigModel}
\end{figure}

Further, we can see that the two-boson sector with the pair-hopping interaction reduces to the effective 1D diagonal Aubry-André or Harper (AAH) single particle model\cite{AubryS80,HarperPG55} with the uniform nearest-neighbor coupling $g$ and modulated commensurate potential $U_m=U\cos(2\pi\alpha m+\phi)$, as shown in Fig.\ref{FigModel} (b). The topological properties of the diagonal AAH model has been identified\cite{LangLJ12PRL, KrausYE12PRL}. It is worth stressing that, although the two-boson Hamiltonian and effective 1D diagonal AAH model have the same energy spectra, there are different bosonic structure properties. The quasiparticles of the two-boson bands sector are the doublon states and the topological doublon edge states have two bosons at the edges. 

\begin{figure}[tb]
\centering
\includegraphics[width=\columnwidth]{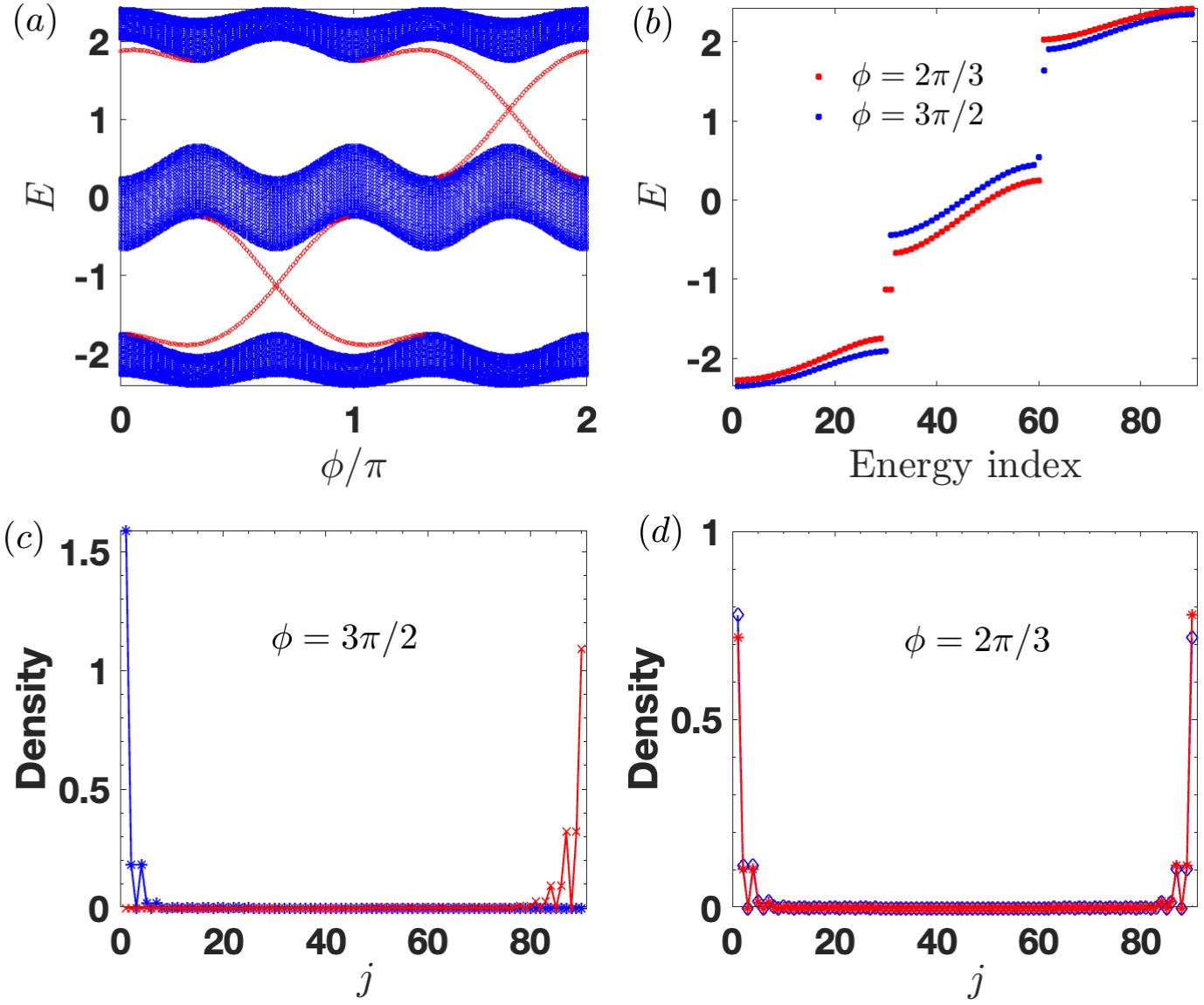}
\caption{$(a)$ The energy spectra as a function of $\phi$ under the OBC. $(b)$ The energy spectra with phase parameter $\phi=3\pi/2$ (blue dots), $2\pi/3$ (red dots) under the OBC. $(c, d)$ The density profiles for the two topological in-gap doublon edge states in the higher energy band gap ($\phi=3\pi/2$) and the lower energy band gap ($\phi=2\pi/3$), respectivly. The other parameters are $N=90, U=1.5, \alpha=1/3$, and $g=1$.}
\label{FigDiagAAH}
\end{figure}

In the following, we analyze the topological features and edge states of this periodically modulated on-site interaction two-boson system. Firstly, we plot the energy spectra [shown in Fig. \ref{FigModel}(c)] calculated by exact diagonalization as a function of $\alpha$ under a periodic boundary condition (PBC) when the system parameters are $U=1.5, \phi=0$, and $g=1$. This is the exact energy spectra of the 1D diagonal AAH model and the 2D Hofstadter model\cite{Hofstadter76PRB}, which are topologically equivalent\cite{KrausYE12PRL}.  For the rational parameter $\alpha$,  we can treat the phase $\phi$ as the momentum of another spatial dimension, the diagonal AAH model could be mapped onto a 2D Hofstadter model with $2\pi\alpha$ magnetic flux per plaquette. Thus, the theoretical tools for analyzing the topological properties of the Hofstadter model could be applied to our system. Here, we should comment that the energy spectra of scattering states (the two bosons propagate independently) have been eliminated. Next, we take the modulated parameter $\alpha=1/3$ as an example \cite{HuangBN24PRL, HuangBN24PS}. Fig. \ref{FigDiagAAH}(a) shows the energy bands as a function of $\phi$ when the system parameters are $N=90, U=1.5$, and $g=1$ under an open boundary condition (OBC). To identify the topological edge states and bulk states, one can use the inverse participation patio (IPR) $\mathrm{IPR_{n}}=\sum_{j}^{N}\left| \psi_{n}(j)\right|^{4}$, where $\psi_{n}$ is the $n$-th normalized eigenstate. For a perfectly extended eigenstate, the $\mathrm{IPR_{n}}$ scales as $1/N$ and vanishes in the thermodynamic limit, while remains a finite value for a localized eigenstate. It is easy to see that there are three Bloch bands for the two-boson bound states. The in-gap states (red dots) are the topological edge states for the two-boson states because the three Bloch bands carry Chern numbers $(1, -2, 1)$, which is calculated in an effective 2D space ($k$, $\phi$) over the Brillouin zone ($0\le k < 2\pi, 0 \le \phi < 2\pi$). Fig. \ref{FigDiagAAH}(b) indicates the energy spectra when phase parameter $\phi=3\pi/2, 2\pi/3$. There are two in-gap doublon states for each $\phi$, where the two in-gap states are degenerate for $\phi=2\pi/3$ because of the inversion symmetry. The density profiles of the in-gap doublon states are illustrated in Fig. \ref{FigDiagAAH}(c) and Fig.\ref{FigDiagAAH}(d), respectively. For the inversion-broken case ($\phi=3\pi/2$), each topological doublon edge state is located in one end region, while the two topological doublon state is located in the two end regions for the inversion symmetry case ($\phi=2\pi/3$). For other modulated period cases, we can use a similar analysis method.

Next, we investigate the nearest-neighbor coupling $J\neq 0$ case. For the scattering bands, we can obtain $E=-4J\cos[(k_1+k_2)/2]\cos[(k_1-k_2)/2]=-2J[\cos k_1+\cos k_2]$, where $k_1$ and $k_2$ are Bloch wavenumbers of the two independent bosons. A powerful approach to the doublon bands and scattering bands of the two-boson sector is provided by the Bethe ansatz technique\cite{GorlachMA17PRA}. However, due to the complexity of the current system, it is difficult to derive the analytical solution for the energy dispersion of the bulk doublon bands ($J\neq 0$ case). In the following, we numerically calculate the energy spectra of doublons and scattering bands using the exact diagonalization.

\begin{figure}[tb]
\centering
\includegraphics[width=\columnwidth]{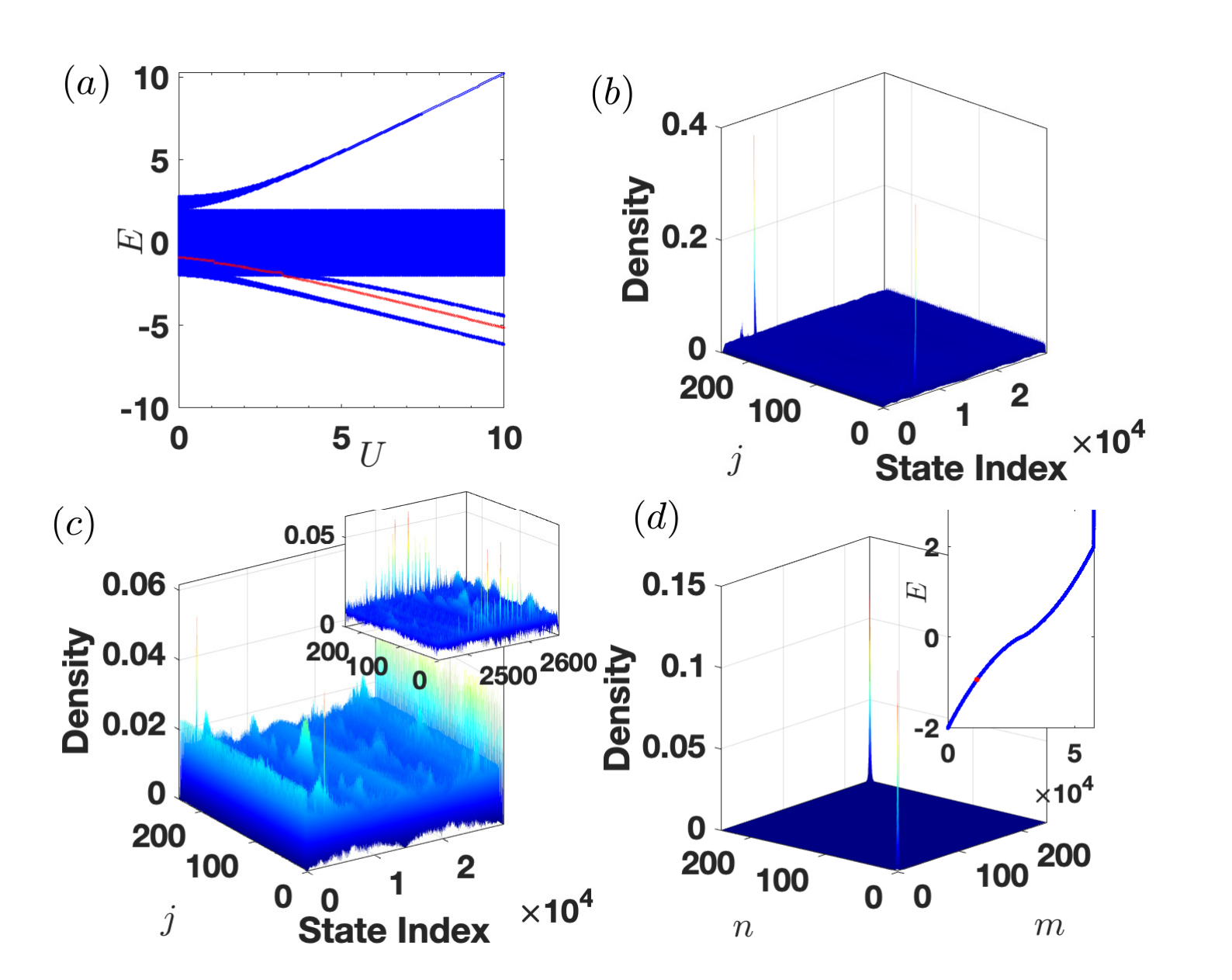}
\caption{$(a)$ The energy spectra as a function of $U$ under the OBC, The red dots represent the topological quasi-degenerate doublon edge states. The other parameters are $N=240, J=0.5, g=1, \alpha=1/3$, and $\phi=2\pi/3$. $(b)$ The density profiles for the doublon bands and scatter bands states at $U=0.2$. (c) The density profiles for the doublon bands and scatter bands states at $U=2$. The inset shows the zoom of the topological quasi-degenerate bound edge states of doublons. $(d)$ The density profile of the topological corner states for the topological metal. The inset illustrates the energy spectra (red dots represent the topological edge states) when the system parameters are $N_m=N_n=240, J=0.5, g=1, U=0.2, \alpha=1/3$, and $\phi=2\pi/3$ under the OBC.}
\label{FigBIC}
\end{figure}

Fig.\ref{FigBIC}(a) shows the evolution of energy spectrum as a function of onsite interaction strength $U$ under the OBC when the system parameters are $N=240, J=0.5, g=1, \alpha=1/3$, and $\phi=2\pi/3$. The red dots stand for the topological quasi-degenerate doublon edge states. From this energy spectra, we can see that for the sufficiently large interaction $U$, the topological two-boson bound edge states are intact and remain in-gap nontrivial states. The 1D interacting boson system is in the topological insulator phase. When the interaction strength is weak, the doublon bulk bands intersect with the scattering bands. When the interaction strength $U=0.2$ and $2$, the density profiles for the doublon bands and scatter bands states (ordered by their respective energy) under OBC are shown in Fig.\ref{FigBIC}(b) and (c), respectively. We can see that the topological  bound states in the continuum (BICs) appear in this topological interacting system, where the bound edge states are embedded in the continuous scattering bands. Unlike the conventional BICs\cite{HsuCW16NRM, ZuoZW23PRB, LinJR25CPB}, the topological bound states in current system are doublon modes localized at the ends\cite{DiLiberto16PRA, GorlachMA17PRA, GorlachMA17PRA2, StepanenkoAA20PRA}. When we use the small coupling $J\neq 0$ term as a perturbation, the effective Hamiltonian of two-particle bound states could be derived by way of perturbation theory\cite{TakahashiM77JPC, BravyiS11AP, QinXZ14PRA, QinXZ18NJP, HuangBN24PRL, HuangBN24PS}, see Appendix \ref{AppendixA} for explicit details. When the coupling $J$ and onsite interaction $U$ are sufficiently weak, the effective Hamiltonian of two-particle bound states is given by
\begin{equation}
H_{eff}=\sum_{j}^{N}(  U_{j}+\frac{2J^{2}}{g})  d_{j}^{\dagger}d_{j}+\sum_{j}^{N-1}(  g+\frac{J^{2}}{g}) (d_{j+1}^{\dagger}d+H.c.)
\end{equation}
where $d_{j}^{\dagger}=b_{j}^{\dagger}b_{j}^{\dagger}/\sqrt{2}$ is the creation operator of the two-boson as a whole at the lattice $j$-site. So, the effective Hamiltonian again becomes the diagonal AAH single particle model. From this diagonal AAH model, we can see that the topological doublon edge states are robust for weak coupling $J$, as shown in Fig.\ref{FigBIC}(b). When the onsite interaction $U$ becomes strong, the topological doublon edge states could become unstable (see Fig.\ref{FigBIC} [c], where the perturbation theory has broken down). It is easy to see that the topological doublon edge states are robust against the weak nearest-neighbor coupling $J$. According to these results, we can infer that when the coupling $J$ becomes further stronger compared with the interaction strength, both the bulk doublon states and topological doublon edge states could become unstable and split into two weakly interacting bosons, where the doublon collapse takes place\cite{GorlachMA17PRA, GorlachMA17PRA2, StepanenkoAA20PRA}. On the other hand, for the mapped 2D single-particle square lattice system, we can infer that the topological insulators and topological metals with one or two corner states could emerge. The density profile of the corner states and energy spectra is shown in Fig.\ref{FigBIC}(d), when the 2D system parameters are $N_m=N_n=240, J=0.5, g=1, U=0.2, \alpha=1/3$, and $\phi=2\pi/3$ under the OBC. Unlike the corner mode states of conventional higher-order topological insulators occupied each corner region\cite{Benalcazar17SCI, Benalcazar17PRB, ZuoZW21JPD}, here the corner states of these topological insulators/metals locate only two diagonal corner regions of the four corners.

\subsection{Periodically modulated pair-hopping interaction}\label{PairHopping}

Next, we investigate the periodically spatial-modulated pair-hopping interaction and uniform on-site interaction case. For the periodically modulated pair-hopping interaction, we choose as $g_j=g[1+\lambda \cos(2\pi\alpha j+\phi)]$, where the parameter $\lambda$ is the modulated strength. Now, the two-boson sector Hamiltonian for the system (the nearest-neighbor coupling $J=0$ case) could be written as
\begin{equation}
g_{m-1}\beta_{m-1,m-1}+g_m\beta_{m+1,m+1}=(E-U)\beta_{m,m}, \label{off-diagAAH}
\end{equation}
with $g_m=g[1+\lambda \cos(2\pi\alpha m+\phi)]$. 

\begin{figure}[tb]
\centering
\includegraphics[width=\columnwidth]{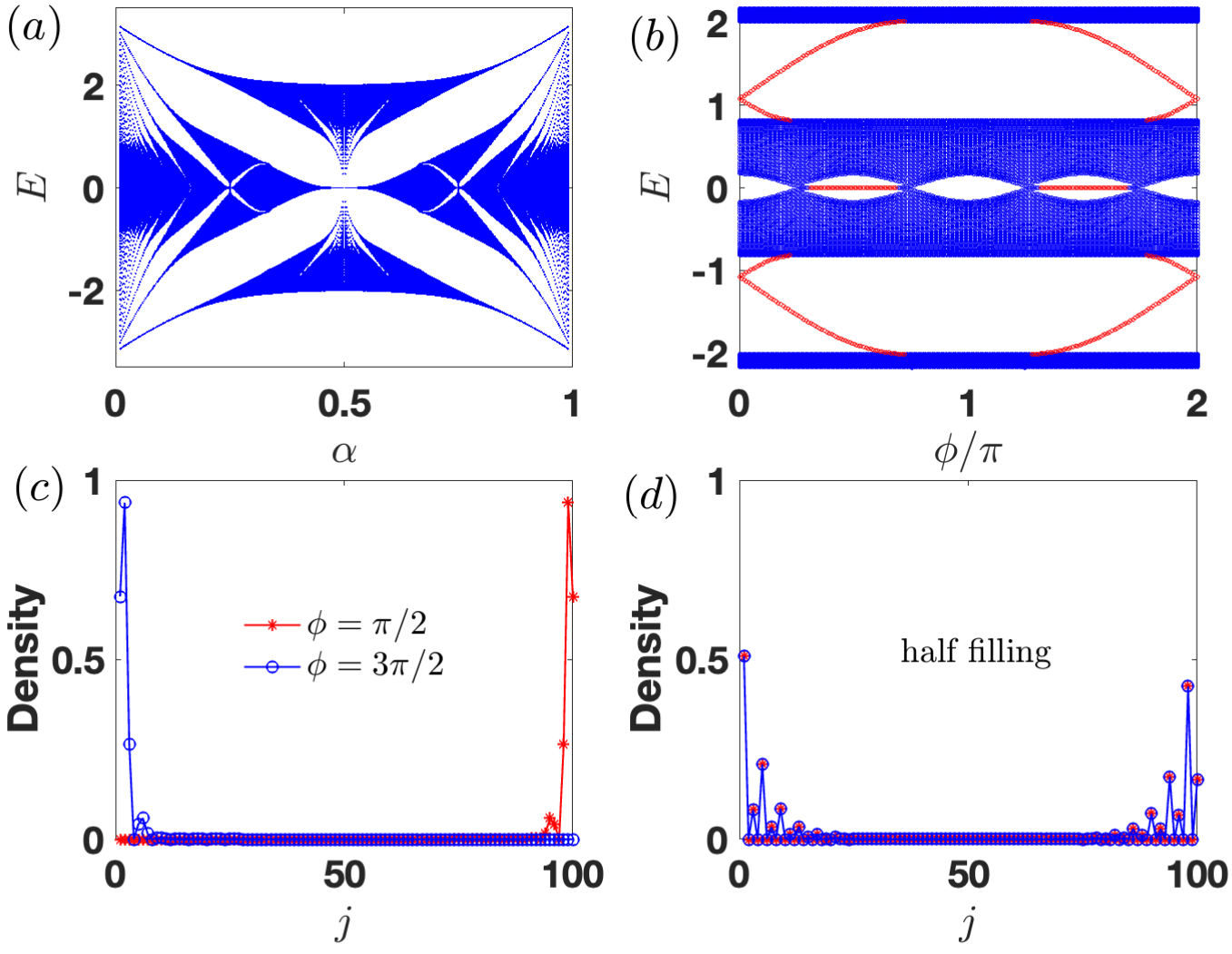}
\caption{$(a)$ The energy spectra as a function of $\alpha$ under the PBC when the system parameters are $g=1, \lambda=0.6$, and $\phi=0$. $(b)$ The energy spectra with $\phi$ under the OBC when $N=100, \lambda=0.6$, and $\alpha=1/4$. $(c)$ The density profiles for the topological in-gap doublon edge states in the lowest energy band gap (red line, $\phi=\pi/2$) and the highest energy band gap: (blue line, $\phi=3\pi/2$). $(d)$ The density profiles for the topological in-gap zero-energy doublon edge states in the center energy band gap. Other parameters are $N=100, g=1, \lambda=0.6, \alpha=1/4$, and $\phi=\pi/2$.}
\label{FigOffDiagAAH}
\end{figure}

The uniform on-site interaction $U$ becomes the on-site uniform potential, thus we can omit this potential. This two-boson sector reduces to the 1D off-diagonal AAH model\cite{GaneshanS13PRL} with the nearest-neighbor coupling $g_m$. So, the topological feature of the off-diagonal AAH model could also be applied to the current system. Fig. \ref{FigOffDiagAAH}(a) shows the energy spectra with $g=1, \lambda=0.6$, and $\phi=0$ when the modulated parameter $\alpha$ varies continuously from 0 to 1, which is also the butterfly-like energy spectra. In the gap regime, we can use the Chern number and winding number to characterize the topological features. It is easy to see that for the modulated parameter $\alpha=1/2$\cite{OlekhnoNA20NTC,OlekhnoNA22PRB}, the off-diagonal AAH model reduces to the SSH model. For the modulated parameter $\alpha=1/4$, the  evolution of the energy spectra with $N=100$, and $\lambda=0.6$ under the OBC is plotted in Fig. \ref{FigOffDiagAAH}(b). In the lowest and highest energy band gaps, there are chiral topological in-gap doublon edge states, because the top and bottom bands have the Chern number 1. The density distributions of the in-gap topological edge states for the two-boson bound pair are illustrated in Fig. \ref{FigOffDiagAAH}(c), where the red (blue) line corresponds to the topological edge states of the doublon in the lowest (highest) energy band gap and $\phi=\pi/2$ $(3\pi/2)$. Next, we analyze the half filling case. There are two topological zero-energy states for the two-boson bound pair because of chiral symmetry. At the half filling, there are two in-gap doublon zero-energy edge states in the $\pi/4<\phi<3\pi/4$ and $5\pi/4<\phi<7\pi/4$ parameter ranges, as shown in Fig. \ref{FigOffDiagAAH}(b). The density profiles for the two zero-energy modes are illustrated in Fig. \ref{FigOffDiagAAH}(d). The topological origin for the two zero-energy edge modes is chiral symmetry and nonzero winding number. 

For the nearest-neighbor coupling $J\neq 0$, we can use the same method for periodically modulated on-site interaction case to unravel the topological features of the  topological insulators and topological metal states of doublons.  When both the pair-hopping interaction and on-site interaction become spatially periodically modulated, the two-boson sector Hamiltonian ($J=0$) could be mapped to the 1D generalized (diagonal and off-diagonal) AAH model\cite{KrausYE12PRL,GaneshanS13PRL}. For the sufficiently small hopping $J$, the second order perturbation theory (see the Appendix \ref{AppendixA}) could also be used to derive the effective Hamiltonian of the doublon subspaces. The similar analysis for the topological states of doublon bands is straightforward. 

\section{Spinless fermion Hubbard case}\label{Fermion}

Armed with the above topological states of the two-boson interacting system, we now transfer to the fermion case. The Hamiltonian for the 1D spinless fermion interacting system with the pair-hopping interaction can be written as
\begin{align}
H_f=&\sum_j^{N-1}[t(c_{j+1}^{\dagger}c_j+H.c.)+V_j n_jn_{j+1}] \notag \\ 
&+\sum_j^{N-2}[g_j c_{j+2}^{\dagger}c_{j+1}^{\dagger}c_{j+1}c_j+H.c.] ,\label{Hf}
\end{align}
where $c_{j}^{\dagger}$ ($c_{j}$) is the spinless fermion creation (annihilation) operator on the lattice site $j$. The first term ($t$) is the nearest-neighbor coupling and the $V_j$ term is the nearest-neighbor interaction. $n_j=c_j^{\dagger}c_j$ is the fermion number operator and the third term ($g_j$) is the pair-hopping interaction from the sites $j, j+1$ to the sites $j+1, j+2$. 

\begin{figure}[tb]
\centering
\includegraphics[width=\columnwidth]{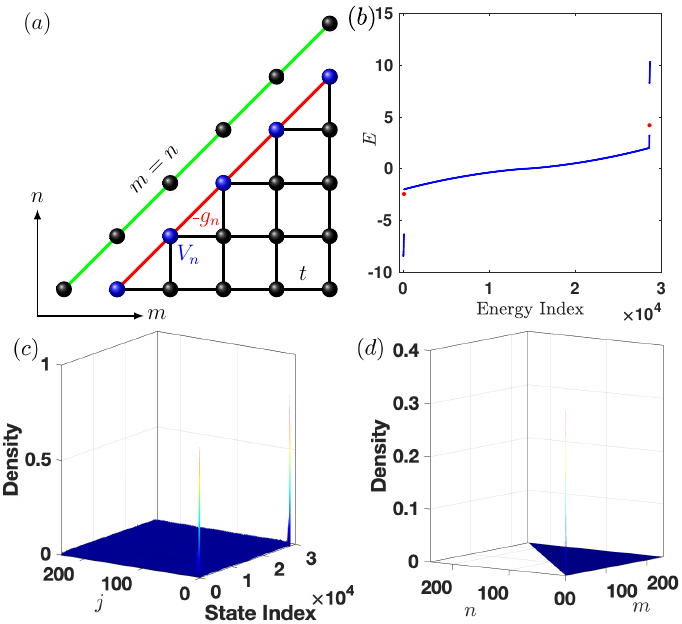}
\caption{$(a)$ Illustration of the mapping onto a 2D single-particle system. $(b)$ The energy spectrum for the doublon and scatter bands under the OBC when the system parameters are $N=240, t=0.5, V=1, g=4.5,\lambda=0.6, \alpha=1/3$, and $\phi=0$. The two red dots are the topological bound states of doublons. $(c)$ The electron densities distribution of the doublons and scatter states. $(d)$ The electron densities distribution of the topological lower-energy corner state in the corresponding 2D $N_m=N_n=239$ system.}
\label{FigFermion}
\end{figure}

Similar to the boson case, we can take the two-particle wave function as $|\psi\rangle=\frac{1}{\sqrt2} \sum_{m, n} \beta_{m,n} c_m^{\dagger} c_n^{\dagger}|0\rangle$. The $\beta_{m,n}=-\beta_{n,m}$ for any indices $m$ and $n$, because of the fermionic anti-symmetry. Unlike the boson case, these $m=n$ lattice sites are a forbidden region due to Pauli's exclusion principle. The mapped 2D square lattices are separated into two independent half-planes. The two states are equivalent and antisymmetric about the $m=n$ axis. Thus, we set the fundamental main of the available states in the lower triangle shape plane ($m>n$). The mapping of this 1D two-fermion problem under the OBC onto the 2D single-particle problem is shown in Fig. \ref{FigFermion}(a). 

If we omit the nearest-neighbor coupling $t$, the system reduces to an effective 1D generalized AAH model with the nearest-neighbor coupling $-g_n$ and superlattice potential $V_n$ and the corresponding linear equation for the 1D generalized AAH model becomes
\begin{equation}
g_{n-1}\beta_{n-1}+g_n\beta_{n+1}=(V_n-E)\beta_{n}. \label{diagFermion}
\end{equation}

Thus, the analysis methods of the AAH model and the above Boson-Hubbard case can also be used to investigate the topological characters for this fermion case. For simplicity, we chose the periodically modulated pair-hopping interaction $g_j=g[1+\lambda \cos(2\pi\alpha j+\phi)]$ and the constant nearest-neighbor interaction $V_j=V$ as an example. When the modulated parameter $\lambda=1/3$, there are three doublon Bloch bands carrying Chern numbers $(1, -2, 1)$. The chiral in-gap topological doublon states would emerge. When the nearest-neighbor coupling $t$ is weak, these topological states are robust. Here, we perform the numerical calculations and show this argument. As an example, Fig. \ref{FigFermion}(b) shows the energy spectra for the doublon and scatter bands with $N=240, t=0.5, V=1, g=4.5,\lambda=0.6, \alpha=1/3$, and $\phi=0$ under the OBC. The red dots display the topological doublon edge states, which are shown in Fig. \ref{FigFermion}(c). These topological doublon edge states are located in one end region because of the inversion symmetry-broken. For the mapped 2D single-particle triangle shape square lattice, the system becomes a topological insulator. The density profile of the topological lower-energy corner state is illustrated in Fig. \ref{FigFermion}(d). As the boson cases, when the parameters of the system obey the inversion symmetry, numerical calculations show that the density profiles of the topological in-gap edge states would be symmetrically located at the two end regions (two corner regions for the mapped 2D system). When the nearest-neighbor coupling $t$ becomes bigger, the topological edge states can enter the scattering bands and turn out to be topological doublon bound states in the continuum and the corresponding 2D system changes into the topological metals. As the nearest-neighbor coupling further increases, the topological edge states would disappear because of the doublon collapse.

\section{Conclusions and Discussions}\label{Conclusions}
In short, the topological properties of two-particle states in the 1D periodically spatial-modulated interacting boson and fermion systems are analyzed. For the two-boson interacting system, the Hamiltonian could be mapped to the diagonal or off-diagonal AAH model in the strongly-correlated limit when the on-site interaction or pair-hopping interaction becomes periodically spatial-modulated type. The doublon bands for the periodically modulated interaction two-boson system exhibit the butterfly-like structure. Compared with the independent-particle  topological insulators, the topological in-gap edge modes in the two-boson interacting topological system are doublon bound states, where the two bosons locate the end sites. When we tune the nearest-neighbor coupling, the doublon bulk bands could intersect with the scattering bands and the topological doublon bound edge states in the continuum emerge. As the nearest-neighbor coupling further increases, the doublon collapse takes place. For the mapped 2D single-particle square system, the topological insulator and topological metal phases with corner states located in the two diagonal corners emerge. For the spinless fermion Hubbard model with pair-hopping interaction, we show the existence of the periodically modulated interaction-induced topological doublon edge states and the realization of the topological insulator 1D interacting system and the topological insulators in the mapped 2D system. The 2D mapped results can be observed in the electrical circuit experiments \cite{OlekhnoNA20NTC,OlekhnoNA22PRB, ZhangWX22NTC, ZhouXQ23PRB, ZhangWX23CP}, and superconducting qubits array\cite{BesedinIS21PRB,TaoZY23arXiv}. Lastly, our findings further demonstrate that the 1D two-particle interacting system and the corresponding 2D system could provide fascinating and promising platforms to study various topological states. 

\section*{Acknowledgements}

This work was supported by the National Natural Science Foundation of China (Grant No. 12074101) and the Natural Science Foundation of Henan (Grant No. 212300410040).

\appendix

\section{Effective Hamiltonian for the doublon states}
\label{AppendixA}
In this Appendix, to understand the effect of the sufficiently small hopping $J$ on the topology of the doublon bands,  we derive the effective Hamiltonian in strong-interaction limit by the perturbation theory\cite{TakahashiM77JPC, BravyiS11AP, QinXZ14PRA, QinXZ18NJP, HuangBN24PRL, HuangBN24PS}. First, we take the hopping term $H_{J}=-J\sum_j^{N-1}(b_{j+1}^{\dagger}b_j+H.c.)$ as a perturbation. Next, we divide the Hilbert space into the subspace $\mathbf{U}$ expanded by the eigenstates $|2\rangle_{j}$ with eigenvalues $E_j=U_j+2g$ and the complement subspace $\mathbf{V}$ expanded by the eigenstates $|1\rangle_{j}|1\rangle_{k}$ ($j\neq k$) with  eigenvalues $E_{j,k}=0$. The projection operators on the two subspace $\mathbf{U}$ and $\mathbf{V}$ can be respectively defined as
\begin{equation}
P=\sum_{j}^{N}|2\rangle_{j}\left\langle 2\right\vert _{j}, \label{Poperator}
\end{equation}
and
\begin{equation}
S =\frac{1}{2}\sum_{j\neq k}^{N}\left(  \frac{1}{E_{j}-E_{j,k}}+\frac{1}{E_{k}-E_{j,k}}\right)  |1\rangle_{j}|1\rangle_{k}\left\langle 1\right\vert_{k}\left\langle 1\right\vert _{j}.\label{Soperator}
\end{equation}

Applying the perturbation theory to the second order, the effective Hamiltonian for the subspace $\mathbf{U}$ is given by
\begin{equation}
H_{eff}=PH_{b}P+PH_{J}SH_{J}P. \label{Heff1}
\end{equation}

Substituting Eqs. \ref{Poperator} and \ref{Soperator} into Eq.\ref{Heff1}, we can obtain the effective Hamiltonian
\begin{widetext}
\begin{equation}
H_{eff} =\sum_{j}^{N}\left[ U_{j}+2J^{2}\left(  \frac{1}{U_{j}+2g}+\frac{1}{U_{j-1}+2g}\right)  \right]  d_{j}^{\dagger}d_{j}+\sum_{j}^{N-1}\left[g+J^{2}\left(  \frac{1}{U_{j}+2g}+\frac{1}{U_{j-1}+2g}\right)  \right]\left(  d_{j+1}^{\dagger}d+H.c.\right)
\end{equation}
\end{widetext}
where $d_{j}^{\dagger}=b_{j}^{\dagger}b_{j}^{\dagger}/\sqrt{2}$ is the creation operator of the two-boson as a whole at the lattice $j$-site. The effective Hamiltonian contains these hopping terms for the two-boson bound states and becomes an effective single-particle Hamiltonian. When the nearest-neighbor hopping $J=0$, the effective Hamiltonian reduce to Eq.\ref{diagAAH}. Under strong pair-hopping interaction ($|J|, |U| \ll |g|$) cases, we can write the effective Hamiltonian as
\begin{equation}
H_{eff}=\sum_{j}^{N}(  U_{j}+\frac{2J^{2}}{g})  d_{j}^{\dagger}d_{j}+\sum_{j}^{N-1}(  g+\frac{J^{2}}{g}) (d_{j+1}^{\dagger}d+H.c.)
\end{equation}
which is the diagonal AAH model. Thus, for sufficiently small hopping $J$, the system maintains the topological properties and topological doublon edge states are robust.

\end{document}